\documentclass{article}

\usepackage{arxiv}

\usepackage[utf8]{inputenc} 
\usepackage[T1]{fontenc}    
\usepackage{hyperref}       
\usepackage[hyperpageref]{backref}
\usepackage{url}            
\usepackage{booktabs}       
\usepackage{amsfonts}       
\usepackage{nicefrac}       
\usepackage{microtype}      
\usepackage{amsmath}
\usepackage{subcaption}
\usepackage{graphicx}

\title{Computation of the phase step between two-step fringe patterns based on Gram--Schmidt algorithm}

\author{
  V\'ictor H. Flores\thanks{viktorhfm@gmail.com} \\
  Department of Computer Science\\
  Centro de Investigaci\'on en Matem\'aticas AC\\
  Guanajuato, MX 36023 \\
  \texttt{victor.flores@cimat.mx} \\
   \And
 Mariano Rivera \\
  Department of Computer Science\\
  Centro de Investigaci\'on en Matem\'aticas AC\\
  Guanajuato, MX 36023 \\
  \texttt{mrivera@cimat.mx} \\
}

\begin{document}
\maketitle

\begin{abstract}
We present the evaluation of a closed form formula for the calculation of the original step between two randomly shifted fringe patterns. Our proposal extends the Gram--Schmidt orthonormalization algorithm for fringe pattern. Experimentally, the phase shift is introduced by a electro--mechanical devices (such as piezoelectric or moving mounts).The estimation of the actual phase step allows us to improve the phase shifting device calibration. The evaluation consists of three cases that represent different pre-normalization processes: First, we evaluate the accuracy of the method in the orthonormalization process by estimating the test step using synthetic normalized fringe patterns with no background,  constant amplitude and different noise levels. Second, we evaluate the formula with a variable amplitude function on the fringe patterns but with no background. Third, we evaluate non-normalized noisy fringe patterns including the comparison of pre-filtering processes such as the Gabor filter banks and the isotropic normalization process, in order to emphasize how they affect in the calculation of the phase step
\end{abstract}


\section{Introduction}
\label{sec:intro}

\noindent The phase shifting method is a well-known technique for the retrieval of the phase from interferometric images \cite{malacara2007optical,servin2014fringe}. Experimentally, these phase shifts are induced using methods such as mirror displacement using a piezoelectric. 

Mathematically, the intensity model of  $n$ phase--shifted interferograms is given by

\begin{equation} \label{eq:image}
I_k(x) = a(x) + b(x)\cos[\phi(x) + \delta_k] + \eta_k(x),
\end{equation}
where $k \in 1,2,\ldots, n$ is the interferogram index, $x=(x_1,x_2)$ is the vector of the pixel coordinates, $a$ and $b$ are the background and the amplitude functions of the interferograms respectively, $\phi$ is the phase to be recovered, $\delta_k$ is the phase step and $\eta_k$ is the noise function. For the case of  two--step algorithms we can assume that $\delta_1 = 0$ and $\delta_2 = \delta$. 

The main issue is the calibration of this components by creating the relation between the applied voltage and the induced phase shift, for this reason, the Gram-Schmidt (GS )orthonormalization algorithm \cite{vargas2012two} is widely used. This algorithm does not require to know the original phase step between the two interferograms since the patterns are orthogonalized in order to compute two interferograms in quadrature (with a phase shift of $\pi/2$ between them). Nevertheless, the GS algorithm relies on the normalization of the fringes through a pre-filtering process \cite{quiroga2003isotropic}, which makes it sensitive to error if the fringes are not correctly normalized.

In this paper we propose a method for estimating the arbitrary  phase step  between two phase shifted fringe patterns. Our method is based on the Gram--Schmidt algorithm and calculates the arbitrary step  with a closed form formula. The proposal consists on evaluating the effects of the variation of background and the amplitude functions as well as the advantages of using the Gabor filter banks\cite{rivera2016two} as pre-filtering process. The computation of the original phase step allows us to improving the phase shifting device calibration (in general, a piezoelectric).

\section{Brief review of Gram--Schmidt orthonormalization for inducing quadrature}
\label{sec:GSA}

\noindent The GS orthonormalization based method, proposed by Vargas \emph{et al.}, calculates the phase between two interferograms with unknown step. 

In this work, we simplify the FP normalization preprocess and assume that just the background intensity variations and noise are removed  \cite{rivera2016two,quiroga2003isotropic,trusiak2015two}. So we start with the process the normalized interferograms: 

\begin{align}\label{eq:imagenorm}
u_1 & = b\, \cos(\phi) \\ 
u_2 & = b \, \cos(\phi + \delta), 
\end{align}
where we omitted the spatial dependency for the vectors $u_1, u_2, b$ and  $\phi$ in order to simplify our notation. According to Ref.  \cite{vargas2012two}, the orthonormalization process consists on 3 steps. First,  $u_1$ is normalized as:

\begin{equation}\label{eq:tildeu1}
\tilde{u}_1 = \frac{ b \cos \phi}{ \| b \cos \phi \|} . 
\end{equation}

Then, $u_2$ is orthogonalized with respect to $\tilde{u}_1$ obtaining its projection as $\hat{u}_{2}$:
\begin{align}
\label{eq:hatu2}
\hat{u}_2  = u_2 - \langle u_2, \tilde{u}_1\rangle \tilde{u}_1 
		 = -b \sin \delta \left[\sin \phi - \kappa \right]
\end{align}
where $\langle \cdot , \cdot \rangle$ represents the inner product and we define  

\begin{equation}
\label{eq:kappa}
\kappa \overset{def}{=} \cos \phi   \frac{\langle b \cos \phi , b \sin \phi\rangle }
    						{\langle b \cos \phi , b \cos \phi \rangle}.
\end{equation}
Since, it is expected  that $\langle b \cos \phi ,  b \sin \phi \rangle << \langle b \cos \phi , b 
\cos \phi \rangle$, then $\kappa$ can be neglected and one has

\begin{equation}
\label{eq:hatu22}
\hat{u}_2  \approx -b \, \sin \delta  \, \sin \phi.  
\end{equation}
Afterwards, $\hat{u}_2$ is normalized:
\begin{equation}\label{eq:tildeu2}
\tilde{u}_2 = -\frac{ b \sin \phi}{ \| b \sin \phi \|} . 
\end{equation}
Finally, by assuming $ \| b \cos \phi \| \approx   \| b \sin \phi \| $ (because the fringes are just shifted but the contribution of valleys and hills remains almost constant),  the wrapped phase is computed with
\begin{equation}
\label{eq:hat_phi}
\hat\phi =  \arctan_2 \left(-\frac{\tilde u_2}{\tilde u_1} \right).
\end{equation}

\section{Calculation of the step}
\label{sec:alpha}

\noindent Herein we introduce the extension to GS algorithm to estimate the actual phase step. For this purpose, we consider that the amplitud term $b(x)$ remains spatially dependent and we estimate the $b$ value using the computed phase $\hat\phi$ as:

\begin{equation}
\label{eq:b}
b(x) = \frac{u_1(x)}{\cos[\hat\phi(x)]}.
\end{equation}
Now,  we substitute \eqref{eq:b} in \eqref{eq:hatu22}, resulting in:
\begin{equation}
\label{eq:hatu23}
\hat{u}_2(x) = - u_1(x) \, \sin \delta   \, \tan [\hat\phi (x)] + \epsilon.
\end{equation}
Afterwards, we use \eqref{eq:hat_phi} and obtain
\begin{equation}
\tilde u_1(x) \hat{u}_2(x)  =    u_1(x) \, \tilde u_2 (x)  \,  \sin \delta  + \epsilon \tilde u_1(x).
\end{equation}
Thus, a $\delta$--map can be computed with
\begin{equation}
\label{eq:delta_x}
 \sin[\delta(x)]  = \frac{ \tilde u_1(x) }{ u_1(x) \tilde u_2(x) } \left[ \hat{u}_2(x) + \epsilon \right]
 \end{equation}
 Then, the phase step  $\delta$ can be estimated by taking the expectation:
  \begin{equation}
\label{eq:delta}
 \delta  =  \arcsin \left( \mathbb{E}_x \left\{ m(x)  \right\}  \right) 
 \end{equation}
where we defined
\begin{equation}
m(x) \overset{def}{=} \frac{\tilde u_1(x) \hat{u}_2(x)}{ u_1(x) \tilde u_2(x)}
\end{equation} 
and we used $\mathbb{E} \left\{r s\right\} =  \mathbb{E} \left\{ r \right\} \mathbb{E} \left\{ s \right\}$ for independent $x$ and $y$; and $\mathbb{E} \left\{ \epsilon \right\} = 0$ by assumption.

In the practice, one can implement the expectation in \eqref{eq:delta} with the mean or,  a more robust estimator, the median. It is also noticeable that the estimation of $\hat\phi(x)$ is not necessary for the calculation of $m(x)$.

If the pre-filtering process removes the amplitude spatial variation, $b=1$, then from the least-squares solution to \eqref{eq:hatu22}, we obtain the closed form formula for $\delta$ is

\begin{equation}
\label{eq:delta2}
\delta = \arcsin \left(- \frac{ \langle  \hat{u}_2(x), \sin \hat\phi(x)  \rangle} {\langle \sin\hat \phi(x), \sin\hat \phi(x)  \rangle } \right).
\end{equation} 

\section{Experiments and results}
\label{sec:exps}
\noindent For the evaluation of the proposed formula \eqref{eq:delta} we will generate 10 sets of synthetic fringe patterns with different noise levels applied to three different cases according to Table \ref{table:cases}.

\begin {table}
\begin{center}
\begin{tabular}{l c l c l c l c |}
Component 	& Case I		&Case II		& Case III		\\
 \hline
$A$ 			& 0 			& 0 			& $a(x)$ 		\\
$B$			& 1			& $b(x)$		& $b(x)$		\\
$\eta(x)$ 		& \checkmark	& \checkmark 	& \checkmark	\\
\hline
\end{tabular}
\caption{Cases of study for the step calculation}
\label{table:cases}
\end{center}
\end{table}

\begin{figure}[h]
	\begin{subfigure}[h]{0.5\textwidth}
		\centering
		\includegraphics[width=.3\textwidth]{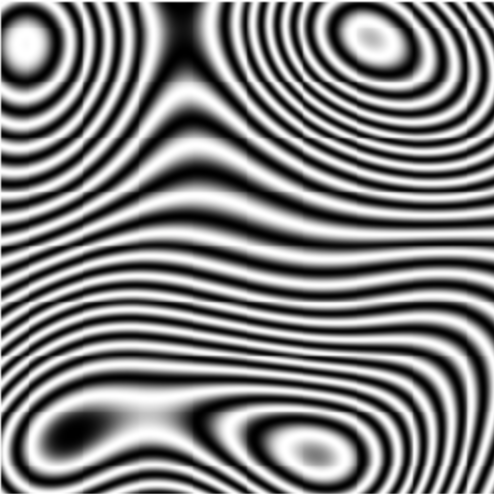}
		\caption{Case I}
		\label{fig:imgc1}
	\end{subfigure}
	\begin{subfigure}[h]{0.5\textwidth}
		\centering
		\includegraphics[width=.3\textwidth]{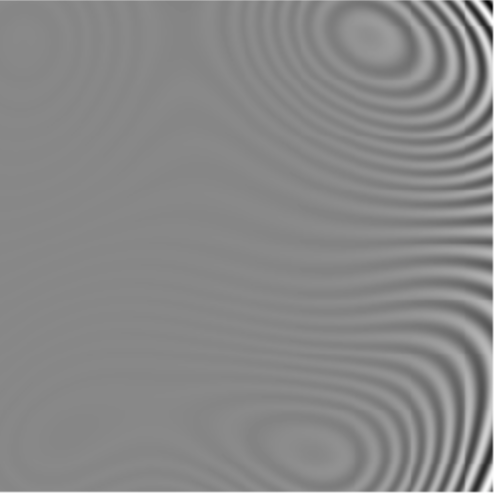}
		\caption{Case II}
		\label{fig:imgc2}
	\end{subfigure}
	\begin{subfigure}[h]{0.5\textwidth}
		\centering
		\includegraphics[width=.3\textwidth]{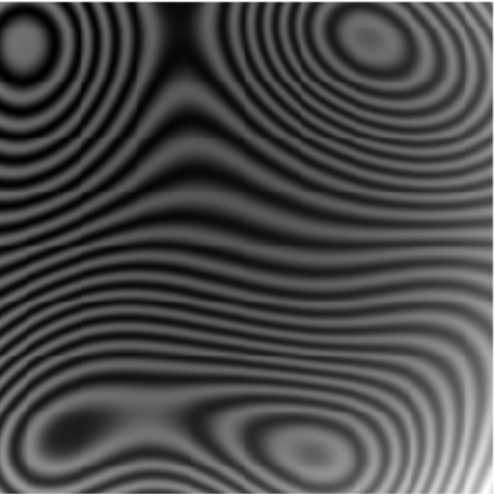}
		\caption{Case III}
		\label{fig:imgc3}
	\end{subfigure}
	
	\begin{subfigure}[h]{0.5\textwidth}
		\centering
		\includegraphics[width=.3\textwidth]{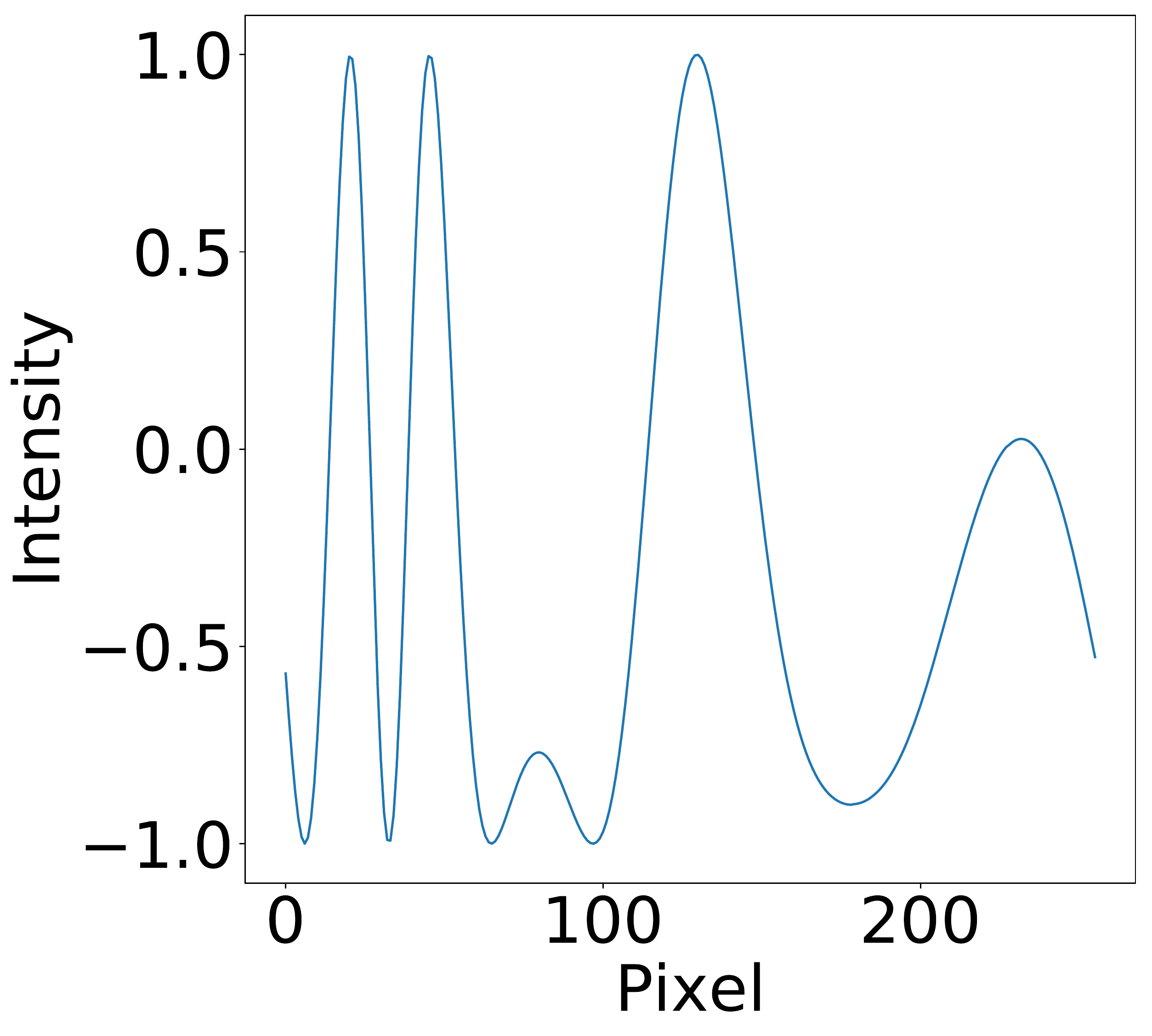}
		\caption{Profile Case I}
		\label{fig:proc1}
	\end{subfigure}
	\begin{subfigure}[h]{0.5\textwidth}
		\centering
		\includegraphics[width=.3\textwidth]{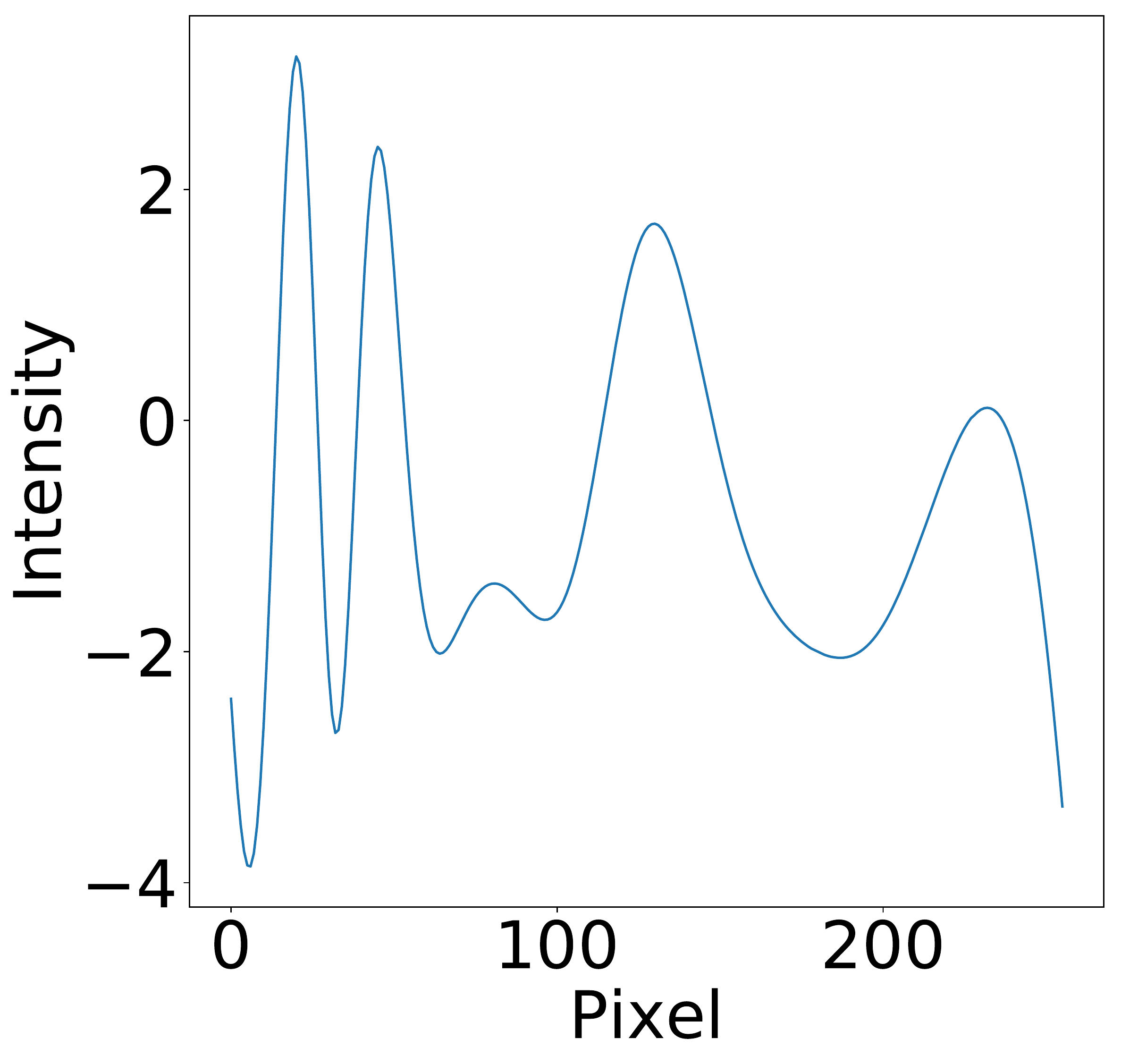}
		\caption{Profile Case II}
		\label{fig:proc2}
	\end{subfigure}
	\begin{subfigure}[h]{0.5\textwidth}
		\centering
		\includegraphics[width=.3\textwidth]{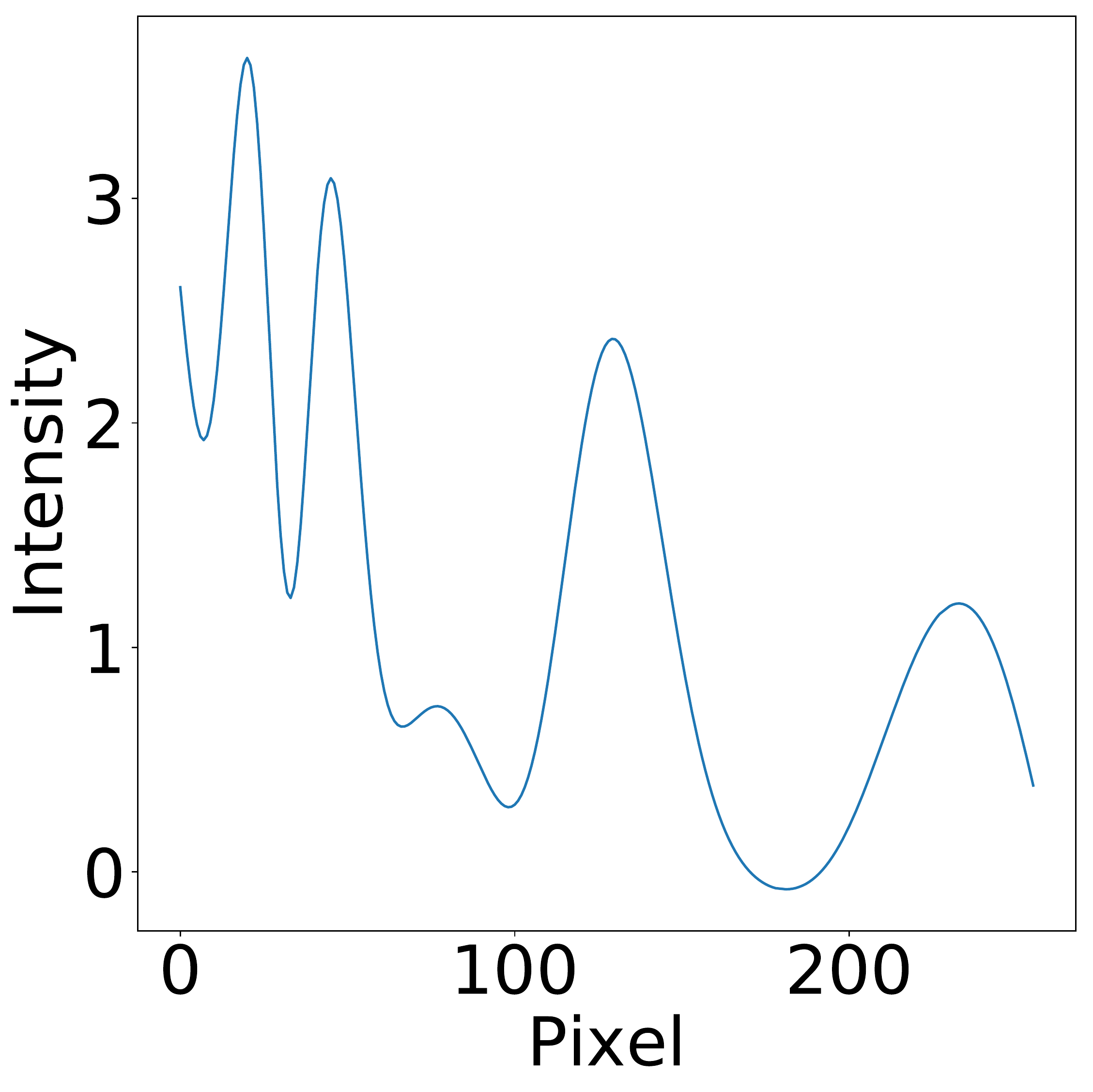}
		\caption{Profile Case III}
		\label{fig:proc3}
	\end{subfigure}
\caption{Synthetic fringe pattern examples and their profiles. (a) and (d) represent Case I, (b) and (e) are Case II and (c) and (f) are for Case III.}
\label{fig:patterns}
\end{figure}

For each case, the actual phase step between the patterns is $\pi/3$ and the noise level variates from $\sigma=0.0$ to $\sigma=1.0$. The sets of images to be used in each case are shown in Figure \ref{fig:patterns} where Figure \ref{fig:imgc1} is a sample of the patterns used in Case I,  \ref{fig:imgc2}  for Case II and  \ref{fig:imgc3} corresponds to Case III. Figures  \ref{fig:proc1}, \ref{fig:proc2} and \ref{fig:proc3} are the profiles of the images of each case. 

\subsection*{Case I}
\begin{figure}[h]
    \centering
    \includegraphics[width=.5\textwidth]{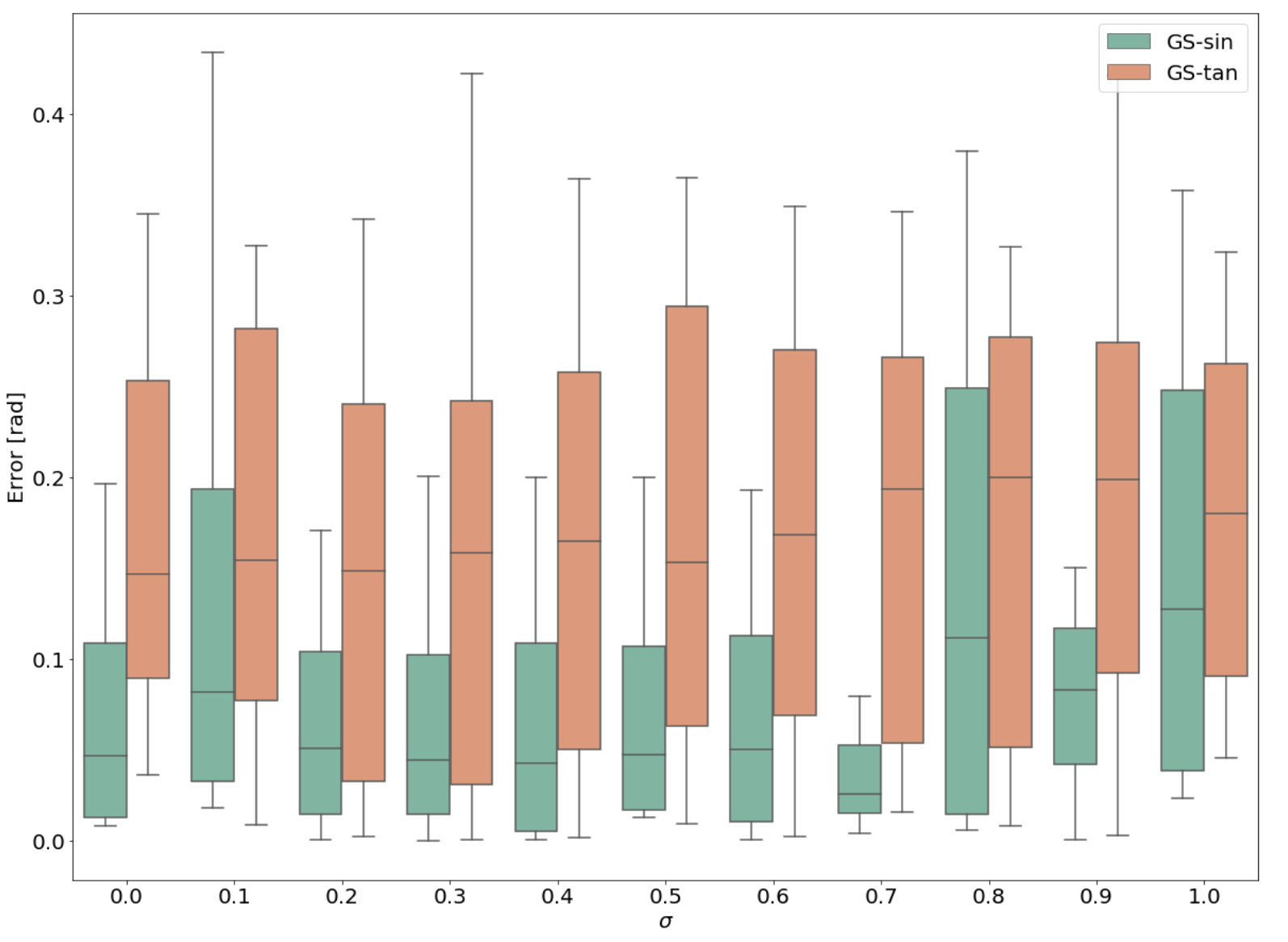}
    \caption{MAE distribution of the phase step estimation using normalized images}
    \label{fig:case1}
\end{figure}

\noindent In order to prove the accuracy of the formula, we estimated the phase step using  \eqref{eq:delta} and and its variation \eqref{eq:delta2} where the amplitude term is constant. The estimation of the step was done using the set of images shown in Figure \ref{fig:imgc1} with ten different noise levels without applying any preprocessing (filter). Figure \ref{fig:case1} shows the results of the estimation where the \textbf{GS-sin} bars correspond to the calculation of the phase step using equation \eqref{eq:delta2} and \textbf{GS-tan} bars correspond to the results of equation \eqref{eq:delta}.

Since the pattern is normalized and the equation \eqref{eq:delta2} is modeled for this special case, it is evident the accuracy on the step calculation. On the other hand, we can observe that the calculation presents stability at certain level of noise, which is acceptable if the goal is to calculate the step.

\subsection*{Case II}
\begin{figure}[h]
    \centering
    \includegraphics[width=.5\textwidth]{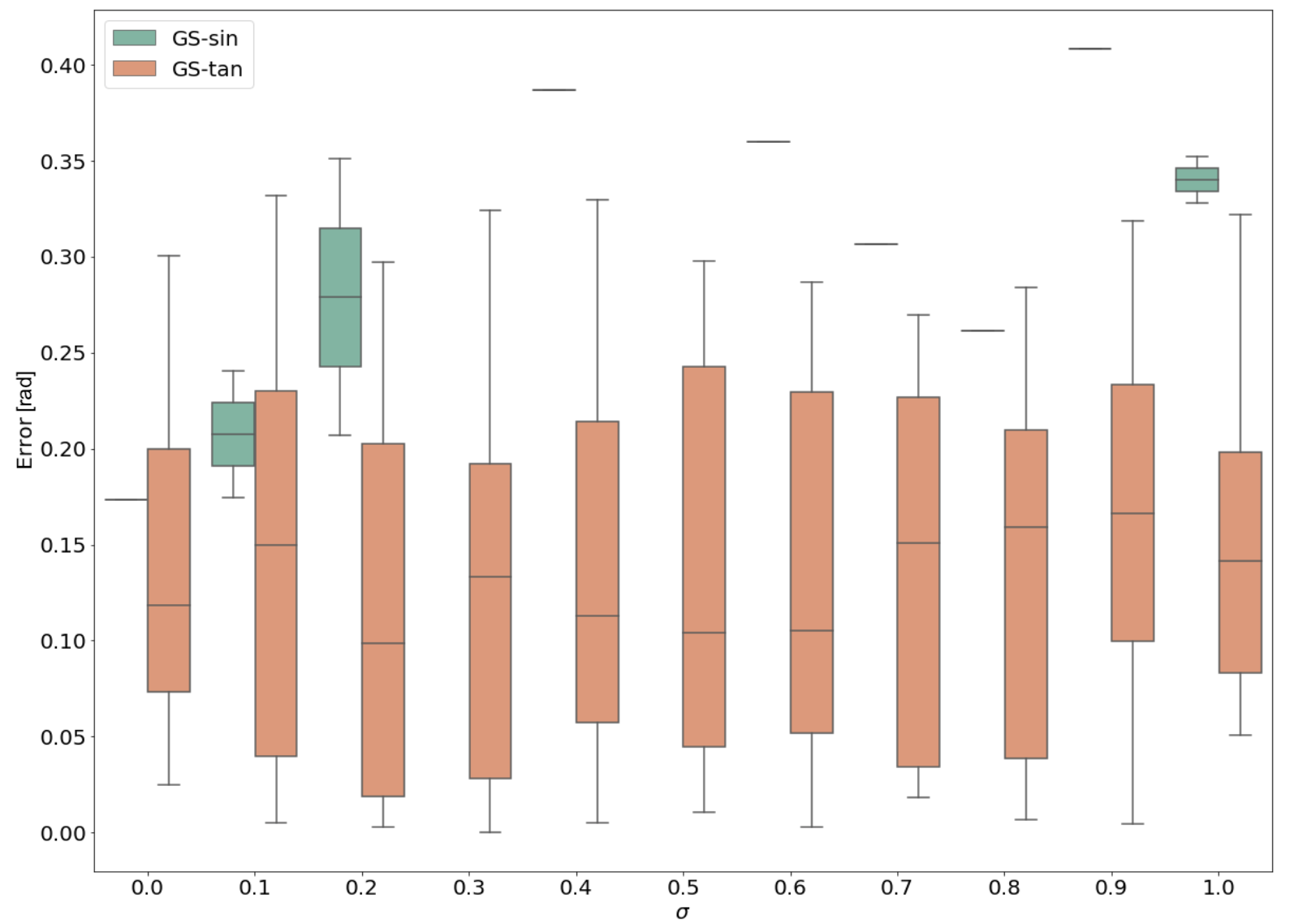}
    \caption{MAE distribution of the phase step estimation using variable amplitude function}
    \label{fig:case2}
\end{figure}

\noindent For this case we did the estimation of the step using the set of images shown in Figure \ref{fig:imgc2} where the amplitude term has spatial dependency. Again, the test was done with ten different noise levels without applying any filter. In Figure \ref{fig:case2}, the results favors to the general formula in  \eqref{eq:delta} because of its robustness to variations in amplitude. In the case of formula in  \eqref{eq:delta2}, there are some levels of noise where it is not obtained a solution since it does not consider the variation of the values of the amplitude function.

\subsection*{Case III}
\begin{figure}[h]
    \centering
    \includegraphics[width=.5\textwidth]{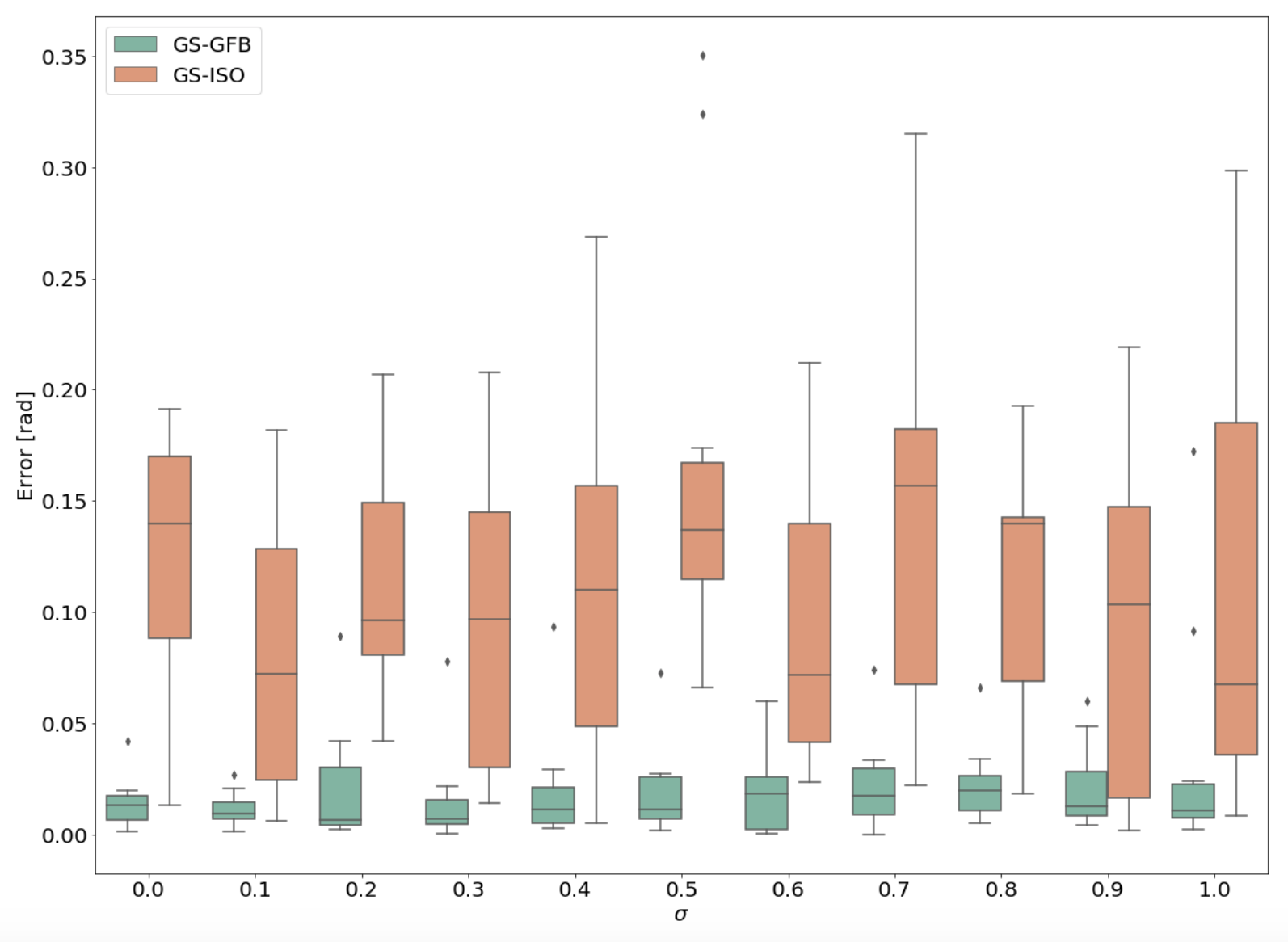}
    \caption{MAE distribution of the phase step estimation using pre-filtered images}
    \label{fig:case3}
\end{figure}

\noindent Finally, the third case consist on using sets of images with variable background and amplitude functions, as well as noise. For this experiment, we compare two different pre-filtering processes: the Gabor Filter Banks (GFB)  \cite{rivera2016two} and the isotropic normalization process \cite{quiroga2003isotropic}; this last one is the originatelly used in the GS algorithm. Figure \ref{fig:case3} shows the error distribution of the phase step estimation.

The calculation was done using \eqref{eq:delta2} since the pre-filtering process delivers normalized images with constant amplitude. It can be seen that the use of GFB reduces significantly the error in the estimation of the phase step.

\section{Discussions \& Conclusions}

\noindent The main contribution of our proposal is that it is possible to calculate the step between two interferograms by using the Gram-Schmidt algorithm. This application is focused on calibrating a phase stepping system based on a piezoelectric in real time. We presented two alternatives for the calculation of the step, a robust one [formula in \eqref{eq:delta}] that only requires to eliminate the background function and the other one [formula in \eqref{eq:delta2}] that considers a normalized pattern with constant amplitude. 

For the case of noisy images with variable background and amplitude, the best option is a pre-filtering process. We presented the comparative of using the Isotropic normalization process, as used in the original algorithm, and the use of GFB. We concluded that the GFB process increased the accuracy of the estimation, nevertheless it is computationally expensive. 

With the obtained results we note that the use of a fast pre-filtering process based on the elimination of the background and using the formula in \eqref{eq:delta},  is enough to estimate the step in real time in order to calibrate the phase shifting system.

\section*{Acknowledgements}
\noindent VHFM thanks Consejo Nacional de Ciencia y Tecnolog\'ia (Conacyt) for the provided postdoctoral grant. This research was supported in part by Conacyt, Mexico (Grant A1-S-43858) and the NVIDIA Academic program.



\end{document}